\documentclass[1p,twocolumn,times,hidelinks]{elsarticle} %5p -> double column journal

\usepackage{lineno,hyperref}
\modulolinenumbers[50]
\let\linenumbers\nolinenumbers\nolinenumbers

\usepackage{xspace}
\xspaceaddexceptions{-}
\usepackage{glossaries}
\usepackage{color}
\usepackage{graphicx}        % standard LaTeX graphics tool
\usepackage{float} % it helps when declaring figures with "H", e.g., "figures[H]"

\usepackage[detect-weight=true, binary-units=true]{siunitx}
\usepackage{amsmath,amssymb,latexsym}
\usepackage{amsfonts}
\usepackage{upgreek}
\usepackage{mathtools}
\usepackage{nicefrac}

\newcommand{\EGnospaces}{Esteban~Garzón}

\newcommand{\EGemail}{esteban.garzon@unical.it}

\newcommand{\RD}{\,Raffaele~De~Rose\xspace}

\newcommand{\FC}{\,Felice~Crupi\xspace}

\newcommand{\AT}{\,Adam~Teman\xspace}

\newcommand{\ML}{\,Marco~Lanuzza\xspace}

\newcommand{\LT}{\,Lionel~Trojman\xspace}

\newcommand{\UNICALfull}{Department of Computer Engineering, Modeling, Electronics and Systems (DIMES), University of Calabria, Rende 87036, Italy}

\newcommand{\BIUwEnICSfull}{Emerging Nanoscaled Integrated Circuits \& Systems (EnICS) Labs, 
Faculty of Engineering, Bar-Ilan University, Ramat-Gan 5290002, Israel}

\newcommand{\ISEPwLISITEfull}{ISEP\,-\,Institut Supérieur d’Électronique de Paris.
10 rue de Vanves, Issy les Moulineaux, 92130, France}

\renewcommand{\eqref}[1]{(\ref{#1})}
\newcommand{\secref}[1]{\mbox{Section~\ref{#1}}}

\newcommand{\figref}[1]{\mbox{Fig.~\ref{#1}}}

\newcommand{\tblref}[1]{\mbox{Table~\ref{#1}}}

\newcommand{\RLtop}{$\text{RL}_\text{T}$\xspace} %{$\text{RL}_{\text{T}}$\xspace}
\newcommand{\RLbottom}{$\text{RL}_\text{B}$\xspace} %{$\text{RL}_{\text{B}}$\xspace}
\newcommand{\tfl}{$t_\text{FL}$\xspace}

\newcommand{\toxt}{$t_\text{OX,T}$\xspace}
\newcommand{\toxb}{$t_\text{OX,B}$\xspace}

\newcommand{\Iwrite}{$I_\text{write}$\xspace}
\newcommand{\Ic}{$I_\text{c}$\xspace}
\newcommand{\Iread}{$I_\text{read}$\xspace}

\newcommand{\var}{$\sigma/\mu$\xspace}

\newcommand{\Vsm}{$V_{\text{SM}}$\xspace}

\newcommand{\tp}{$t_{\text{p}}$\xspace}
\newcommand{\tread}{$t_{\text{read}}$\xspace}

\newcommand{\Rapt}{$R_{\text{AP,T}}$\xspace}
\newcommand{\Rapb}{$R_{\text{AP,B}}$\xspace}
\newcommand{\Rpt}{$R_{\text{P,T}}$\xspace}
\newcommand{\Rpb}{$R_{\text{P,B}}$\xspace}

\newcommand{\RL}{$R_{\text{L}}$\xspace}
\newcommand{\RH}{$R_{\text{H}}$\xspace}

\newcommand{\Ms}{$M_{\text{S}}$\xspace}

\newcommand{\Ki}{$K_{\text{i}}$\xspace}

\newcommand{\Ids}{$I_{\text{DS}}$\xspace}
\newcommand{\Vds}{$V_{\text{DS}}$\xspace}
\newcommand{\Vgs}{$V_{\text{GS}}$\xspace}
\newcommand{\Vbs}{$V_{\text{BS}}$\xspace}

\newcommand{\MTJconductance}{G(\theta)=G_{T}[1+P^{2}\cos(\theta)]+G_{SI}}  % MTJ Conductance
\newcommand{\SpinPolarizationFactor}{P(T)=P(0)(1-\beta T^{3/2})}   % Spin polarization factor
\newcommand{\TMRzeroBias}{\mathit{TMR}(T,0)=\frac{2P^{2}}{(1-P^{2})+G_{SI}/G_{T}}}   % Temperature-dependent TMR @zero bias
\newcommand{\TMRvoltageDependent}{\mathit{TMR}(T,V)=\frac{\mathit{TMR}(T,0)}{1+(\frac{V_\text{OX}}{V_{H}})^2}}    % Voltage-dependent TMR

\newcommand{\SaturationMagnetization}{M_{S}(T)=M_{S}(0)\left[1-\left(\frac{T}{T^{*}}\right)^{3/2}\right]} % saturation magnetization 
\newcommand{\InterfacialPerpendicularAnisotropy}{K_{i}(T)=K_{i}(0)\left(\frac{M_{S}(T)}{M_{S}(0)}\right)^{2.18}}  % interfacial perpendicular anisotropy
\newcommand{\EffectiveAnisotropyField}{H_\mathit{k,eff}=\frac{2K_{i}}{t_\text{FL}\mu_{0}M_{S}} - (N_{Z}-N_{X,Y})M_S} % effective anisotropy field 
\newcommand{\deltaDW}{ 
    \begin{cases}
      \Delta = \cfrac{K_{EFF} V_{FL}}{k_B T} & \left(D \leq D_W\right)\\
      \Delta = \cfrac{\pi^3 A_{ex} t_{FL}}{4k_B T} & \left(D > D_W\right)\\
    \end{cases} 
    } % Fitting parameter beta

\newcommand{\KEFFa}{K_{EFF} = \cfrac{K_i}{t_{FL}} - \cfrac{\mu_0 M_{S}^2}{2}\left(N_Z - N_{X,Y}\right)}

\newcommand{\MTJCriticalSwitchingCurrentVa}{I_{c}=\frac{\alpha e \gamma_{0} \mu_{0} H_\mathit{k,eff} M_{S}~ V_\text{FL}}{\mu_{B}~g_\text{STT}}} % Critical switching current 

\newcommand{\gDMTJ}{g_{STT}=\frac{4P}{1-P^{4}}}

\newcommand{\uA}{\,\si{\micro\ampere}\xspace}

\newcommand{\V}{\,\si{\volt}\xspace}
\newcommand{\mV}{\,\si{\milli\volt}\xspace}

\newcommand{\Kohms}{\,\si{\kilo\ohm}\xspace}
\newcommand{\ohms}{\,\si{\ohm}\xspace}

\newcommand{\J}{\,\si{\joule}\xspace}

\newcommand{\fJ}{\,\si{\femto\joule}\xspace}

\newcommand{\msquared}{\,\si{\meter\squared}\xspace}

\newcommand{\nm}{\,\si{\nano\meter}\xspace}

\newcommand{\um}{\,\si{\micro\meter}\xspace}
\newcommand{\umsquared}{\,\si{\micro\meter\squared}\xspace}

\newcommand{\ns}{\,\si{\nano\second}\xspace}

\newcommand{\bit}{\,\si{\bit}\xspace}

\newcommand{\KB}{\,\si{\kilo\byte}\xspace}

\newcommand{\MB}{\,\si{\mega\byte}\xspace}

\newcommand{\pW}{\,\si{\pico\watt}\xspace}

\newcommand{\K}{\,\si{\kelvin}\xspace}

\newcommand{\T}{\,\si{\tesla}\xspace}

\newcommand{\X}{\,$\times$\xspace}    % times -> x
    
\newacronym{ASIen}{ASI}{Italian Space Agency}
    
\newacronym{ASIit}{ASI}{Agenzia Spaziale Italiana}
    
\newacronym{UNICALen}{UNICAL}{University of Calabria}
    
\newacronym{UNICALit}{UNICAL}{Università della Calabria}
    
\newacronym{DIMESen}{DIMES}{Department of Computer Engineering, Modeling, Electronics and Systems}
    
\newacronym{DIMESit}{DIMES}{Dipartimento di Ingegneria Informatica, Modellistica, Elettronica e Sistemistica}
    
\newacronym{USFQes}{USFQ}{Universidad San Francisco de Quito}
         
\newacronym{BIUen}{BIU}{Bar-Ilan University}
     
\newacronym{tsmc}{TSMC}{Taiwan Semiconductor Manufacturing Company}
     
\newacronym{beol}{BEOL}{back-end-of-line}

\newacronym{feol}{FEOL}{front-end-of-line}

\newacronym{wordl}{WL}{wordline}
    
\newacronym{bitl}{BL}{bitline}
    
\newacronym{sourcel}{SL}{sourceline}
    
\newacronym[longplural={nanowires}]{nw}{NW}{nanowire}

\newacronym{ptm}{PTM}{predictive technology model}

\newacronym{pdk}{PDK}{process design kit}
    \newcommand{\pdk}{\gls{pdk}\xspace}

\newacronym[longplural={systems-on-chip}]{soc}{SoC}{system-on-chip}

\newacronym{ai}{AI}{artificial-intelligence}

\newacronym{iot}{IoT}{internet-of-things}

\newacronym{mc}{MC}{Monte Carlo}
    \newcommand{\mc}{\gls{mc}\xspace} 
\newacronym{cvs}{CVS}{conventional voltage sensing}

\newacronym{epi}{EPI}{energy per instruction}

\newacronym{ips}{IPS}{instructions per second}

\newacronym{mep}{MEP}{minimum energy point}

\newacronym{lrs}{LRS}{low resistance state}

\newacronym{hrs}{HRS}{high resistance state}

\newacronym{mim}{MIM}{metal-insulator-metal}

\newacronym[longplural={phase-change memories}]{pcm}{PCM}{phase-change memory}

\newacronym[longplural={resistive RAMs}]{rram}{RRAM}{resistive RAM}

\newacronym[longplural={spin-transfer torque magnetic RAMs}]{sttmram}{STT-MRAM}{spin-transfer torque magnetic RAM}
    \newcommand{\sttmram}{\gls{sttmram}\xspace} 
     
    \newcommand{\sttmrams}{\glspl{sttmram}\xspace}
    \newcommand{\Sttmrams}{\Glspl{sttmram}\xspace}     
\newacronym{euv}{EUV}{extreme ultra-violet}
     
\newacronym[longplural={Gain-Cell embedded DRAMs}]{gcedram}{GC-eDRAM}{Gain-Cell embedded DRAM}

\newacronym{sixt}{6T}{6-transistor}
    
\newacronym{eflash}{eFlash}{embedded Flash}

\newacronym[longplural={multi-level cells}]{mlc}{MLC}{multi-level cell}

\newacronym[longplural={Storage Class Memories}]{scm}{SCM}{Storage Class Memory}

\newacronym{ddr}{DDR}{dual-data rate}

\newacronym[longplural={graphic processing units}]{gpu}{GPU}{graphic processing unit}

\newacronym[longplural={central processing units}]{cpu}{CPU}{central processing unit}

\newacronym{sata}{SATA}{Serial Advanced Technology Attachment}
    
\newacronym{nvme}{NVMe}{Non-Volatile Memory Express}
    
\newacronym{nvm}{NVM}{NVM}
        
\newacronym{pcie}{PCIe}{Peripheral Component Interconnect Express}
     
\newacronym[longplural={hard-Disk drives}]{hdd}{HDD}{hard-Disk drive}

\newacronym[longplural={solid-State drives}]{ssd}{SSD}{solid-State drive}

\newacronym[longplural={high-bandwidth memories}]{hbm}{HBM}{high-bandwidth memory}

\newacronym[longplural={dual-inline memory modules}]{dimm}{DIMM}{dual-inline memory module}

\newacronym[longplural={static random-access memories}]{sram}{SRAM}{static random-access memory}

\newacronym[longplural={embedded DRAMs}]{edram}{eDRAM}{embedded DRAM}

\newacronym[longplural={dynamic random-access memories}]{dram}{DRAM}{dynamic random-access memory}

\newacronym[longplural={six-transistor static random access memories}]{sixtsram}{6T-SRAM}{six-transistor static random access memory}
    \newcommand{\sixtsram}{\gls{sixtsram}\xspace}

\newacronym[longplural={magnetic random-access memories}]{mram}{MRAM}{magnetic random-access memory}

\newacronym{bc}{BC}{bitcell}

\newacronym{bl}{BL}{bitline}

\newacronym{sln}{SL}{sourceline}

\newacronym{wl}{WL}{wordline}

\newacronym{nc}{NC}{number of cycles at endurance failure}
     
\newacronym{llgs}{LLGS}{Landau-Lifshitz-Gilbert-Slonczewski}
     
\newacronym{stt}{STT}{spin-transfer torque}

\newacronym{pma}{PMA}{perpendicular magnetic anisotropy}

\newacronym{ima}{IMA}{in-plane magnetic anisotropy}

\newacronym{mtj}{MTJ}{magnetic tunnel junction}

\newacronym{smtj}{SMTJ}{single-barrier MTJ}
    \newcommand{\smtj}{\gls{smtj}\xspace}

\newacronym{dmtj}{DMTJ}{double-barrier MTJ}
    \newcommand{\dmtj}{\gls{dmtj}\xspace} 
     
    \newcommand{\dmtjs}{\glspl{dmtj}\xspace}
    
\newacronym{mr}{MR}{magnetoresistance}

\newacronym{tmr}{TMR}{tunnel magnetoresistance}

\newacronym{gmr}{GMR}{giant magnetoresistance}

\newacronym{wer}{WER}{write error rate}

\newacronym{rdr}{RDR}{read disturbance rate}
    \newcommand{\rdr}{\gls{rdr}\xspace} 
     
\newacronym{rfr}{RFR}{retention failure rate}

\newacronym{rer}{RER}{read error rate}

\newacronym{fl}{FL}{free layer}
    \newcommand{\fl}{\gls{fl}\xspace}

\newacronym[longplural={reference layers}]{rl}{RL}{reference layer}

\newacronym{fm}{FM}{ferromagnetic}
    \newcommand{\fm}{\gls{fm}\xspace}

\usepackage{booktabs}
\usepackage{threeparttable}

\usepackage{float}
\usepackage{multirow}
\usepackage{subfiles} % Best loaded last in the preamble
\usepackage{soul}
\sethlcolor{white} %--> to get rid of the yellow hl, un-comment this
\makeatletter
\def\ps@pprintTitle{%
  \let\@oddhead\@empty
  \let\@evenhead\@empty
  \let\@oddfoot\@empty
  \let\@evenfoot\@oddfoot
}
\makeatother
\begin{document}

\begin{frontmatter}

\title{Adjusting Thermal Stability in Double-Barrier MTJ for Energy Improvement in Cryogenic STT-MRAMs}

\author[A,B]{\EGnospaces\corref{correspondingauthor}}
\ead{\EGemail}
\author[A]{\RD}
\author[A]{\FC}
\author[C]{\LT}
\author[B]{\AT}
\author[A]{\ML} 

\cortext[correspondingauthor]{Corresponding author}

\address[A]{\UNICALfull}
\address[B]{\BIUwEnICSfull}
\address[C]{\ISEPwLISITEfull}


\begin{abstract}
This paper investigates the impact of thermal stability relaxation in double-barrier magnetic tunnel junctions (DMTJs) for energy-efficient spin-transfer torque magnetic random access memories (STT-MRAMs) operating at the liquid nitrogen boiling point (77\K).
Our study is carried out through a macrospin-based Verilog-A compact model of DMTJ, along with a 65\nm commercial process design kit (PDK) calibrated down to 77\K under silicon measurements.
Comprehensive bitcell-level electrical characterization is used to estimate the energy/latency per operation and leakage power at the memory architecture-level.
As a main result of our analysis, we show that energy-efficient small-to-large embedded memories can be obtained by significantly relaxing the non-volatility requirement of DMTJ devices at room temperature (i.e., by reducing the cross-section area), while maintaining the typical 10-years retention time at cryogenic temperatures.
This makes DMTJ-based STT-MRAM operating at 77\K more energy-efficient than  six-transistors static random-access memory (6T-SRAM) under both read and write accesses (--56\% and --37\% on average, respectively).
Obtained results thus prove that DMTJ-based STT-MRAM with relaxed retention time is a promising alternative for the realization of reliable and energy-efficient embedded memories operating at cryogenic temperatures.
\end{abstract}

\begin{keyword}
Double-barrier magnetic tunnel junction (DMTJ), STT-MRAM, Cryogenic electronics, Cryogenic cache, Thermal stability relaxation, 77\K
\end{keyword}

\end{frontmatter}

\linenumbers

\section{Introduction}
\label{sec:intro}
\Sttmrams have been gaining interest for cryogenic computing electronics due to their reduced area footprint, improved readout capabilities, and increased data retention time~\cite{irds2020,garzon2022embedded,chiang2021cold}.
In this regard, recent studies have focused on \sttmram operating down to the liquid nitrogen boiling point (77\K), showing that it is a viable energy-efficient alternative to \sixtsram for medium-to-large embedded memories under cryogenic operation~\cite{chiang2021cold,garzon2021cryoSB,garzon2021cryoDB}. %garzon2021relaxing 
In view of the above benefits, \sttmrams are identified as potential candidates for future cryogenic computing applications.
However, to better support cryogenic computing, energy improved  memories with increased storage density are particularly sought after.
\hl{To deal with this, tuning thermal stability of magnetic tunnel junction (MTJ) devices has been recently considered as a promising approach~\mbox{\cite{chiang2021cold,garzon2021relaxing}}}.

In the above context, this work aims to demonstrate a potential solution to build reliable, energy-efficient, and high-density \sttmrams operating at cryogenic conditions (77\K).
This is achieved by exploiting the reduced switching currents of the \dmtj  as compared to its conventional \smtj counterpart, along with the concept of relaxing its non-volatility requirement at room temperature (i.e., by reducing its cross-section area), while maintaining the typical 10-years retention time at cryogenic temperatures. 

\begin{figure*}[h!] % [b]-> bottom, [t]->top, [H]->Here! ([h!] should do a better job), {figure*}->float
    \centering 
    \includegraphics[width=1.0\columnwidth]{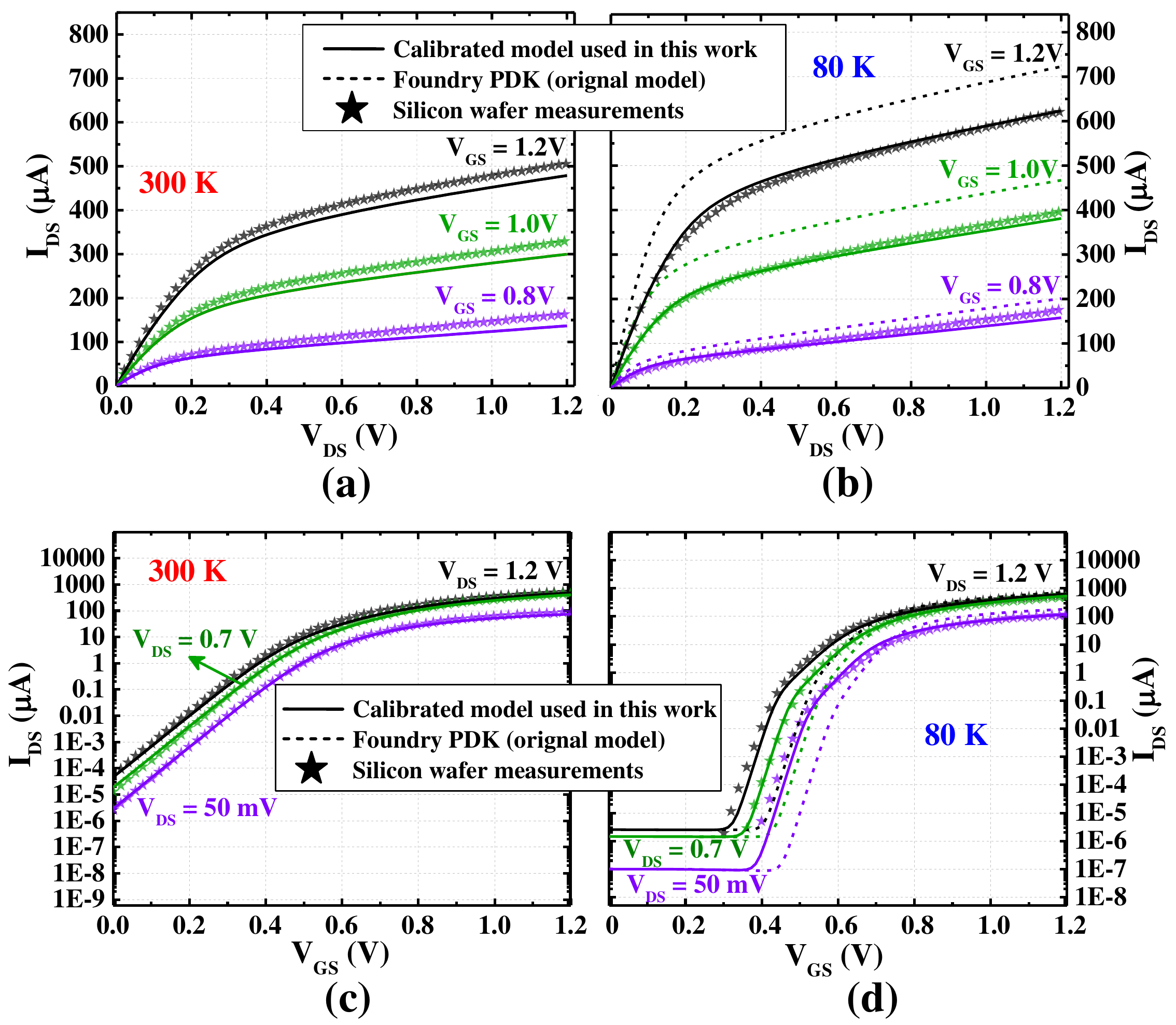}
    \vspace{-3mm}
    \caption{CMOS current-voltage characteristics at room and cryogenic temperatures: (a)-(b) \Ids vs. \Vds for different \Vgs at 300\K and 80\K, (c)-(d) \Ids vs. \Vgs for different \Vds at 300\K and 80\K.
    Data refers to an nMOS transistor with \emph{L} = 60\nm and \emph{W} = 1\um at body-source voltage \Vbs = 0\V.}
    \label{fig:MOS_idsvgs}
\end{figure*}

Our study is carried out using a cross-layer simulation framework, which spans from the device-level up to the memory architecture-level, passing through bitcell-level electrical simulations~\cite{garzon2021cryoSB,garzon2021cryoDB,garzon2021relaxing,garzon2022voltage}.
DMTJs and transistors are respectively modeled using a macrospin-based Verilog-A compact model~\cite{de2019compact} and a commercial 65\nm CMOS technology that was fully characterized down to 77\K.
Bitcell-level simulations under cryogenic operation benchmark a 13\nm-DMTJ based bitcell with relaxed non-volatility at room temperature and 10-years retention at 77\K against its 40\nm-DMTJ based counterpart providing 10-years retention at 300\K.
Our comparative analysis is extended to the architecture-level by considering cache memory sizes ranging from 64\KB up to 2\MB.
\hl{In our study, we also considered the conventional 6T-SRAM, whose electrical characteristics were extracted from simulations at 77\K.
Overall, while cryogenic temperatures enable STT-MRAM improvements in terms of data retention and memory window as given respectively by larger thermal stability and higher tunneling magnetoresistance (TMR) ratio, the 6T-SRAM benefits from significant static power savings.}
As the main outcome of our work, 
\hl{we show that, over the considered memory capacity range and thanks to the non-volatility of the DMTJ properly tuned for cryogenic operation, the DMTJ-based STT-MRAM operating at 77\,K proves to be more energy-efficient than conventional 6T-SRAM under both read and write operations by 56\% and 37\% (on average), respectively.}

\hl{This work details from device-level up to architecture-level the benefits of adjusting thermal stability for friendly cryogenic STT-MRAMs.
It is worth mentioning that this paper extends the study presented by the authors in~\mbox{\cite{garzon2021relaxing}} by detailing the device-level modeling and analysis for both silicon-calibrated transistor models and DMTJ devices under cryogenic operation, while also describing the fine-grained calibration procedure for cryogenic memory architecture-level simulations.
As further difference over~\mbox{\cite{garzon2021relaxing}}, reported simulation results account for the effect of the domain-wall in DMTJs, which has been included in the adopted macrospin-based model according to the approach described in~\mbox{\cite{wang2019compact}}}

The rest of the paper is organized as follows.
\secref{sec:DeviceLevel} describes device-level characteristics at cryogenic temperatures.
\secref{sec:Analysis77K} discusses bitcell- and architecture-level simulation results of cryogenic DMTJ-based STT-MRAM, while discussing the adopted simulation approach.
Finally, \secref{sec:Conclusions} concludes this work.

\section{Device-level modeling and analysis at cryogenic temperatures}
\label{sec:DeviceLevel}
\subsection{CMOS devices under cryogenic operation}
Unlike previous cryogenic computing studies~\cite{shin2014low,zhao2014modeling} that extend the MOSFET BSIM4 model for the target cryogenic temperature using data reported in literature, our work relies on a commercial 65\nm \pdk whose BSIM4.7-based \hl{equations were calibrated with silicon measurements at cryogenic temperatures.
Such calibration has concerned different temperature-dependent model parameters, e.g., leakage, threshold voltage, mobility, body factor, series resistances, stress effects, etc.
Likewise, the adopted cryogenic-aware PDK also provides statistical models}
to consider the impact of manufacturing uncertainty on transistor characteristics.

\figref{fig:MOS_idsvgs}(a)-(d) show the comparison among the original foundry \pdk model, the calibrated model, and silicon measurements in terms of current-voltage characteristics at two different temperatures (300\K and 80\K) for an nMOS transistor with channel length \emph{L} = 60\nm and channel width \emph{W} = 1\um.
From \figref{fig:MOS_idsvgs}(a) and (c), simulation results of the original model and calibrated model are roughly the same at 300\K, both well-tracking experimental data.
However, as the temperature goes down to 80\K, the calibrated model tracks silicon measurements much more accurately, as shown in \figref{fig:MOS_idsvgs}(b) and (d).
To give a reference, the average error of the original model with respect to experimental data at 80\K is more than 18\% and 70\% in terms of drain-source current (\Ids) versus drain-source voltage (\Vds) for gate-source voltage \Vgs = 1.2\V and \Ids versus \Vgs for \Vds = 1.2\V, respectively.
Conversely, for the same bias conditions, the average error achieved by the calibrated model at 80\K is respectively less than 2\% and 10\% as compared to silicon measurements.

Regarding the impact of cryogenic operation on CMOS characteristics, the main effects are related to increase the charge carrier mobility, saturation velocity, and threshold voltage, as well as an improvement in sub-threshold slope~\cite{balestra2001device,trojman2008mobility,zhao2014modeling,8329135}. 
Overall, when cooling down to cryogenic temperatures, CMOS-based circuits can benefit from increased ON/OFF current ratio as shown in Fig. 1, thus resulting in better performance and reduced standby power.
Indeed, from \figref{fig:MOS_idsvgs}(a)-(d), the ON current (i.e., \Ids for \Vds = \Vgs = 1.2\V) increases by about 30\% when reducing the temperature from 300\K down to 80\K, whereas the OFF current (i.e., \Ids for \Vds = 1.2\V and \Vgs = 0\V) decreases about 1.3 orders of magnitude.

Note that both bitcell- and architecture-level analyses, reported below, are based on the aforementioned 
\hl{cryogenic-aware PDK calibrated on real silicon measurements. Therefore, when compared to the use of predictive models or estimated extrapolations from PDK models calibrated within the typical commercial temperature range, this ensures more reliable simulation-based results at the operating point of 77\,K.}

\subsection{DMTJ devices under cryogenic operation}
\begin{figure}[!b] % [b]-> bottom, [t]->top, [H]->Here! ([h!] should do a better job), {figure*}->float
    \centering
    \includegraphics[width=1\columnwidth]{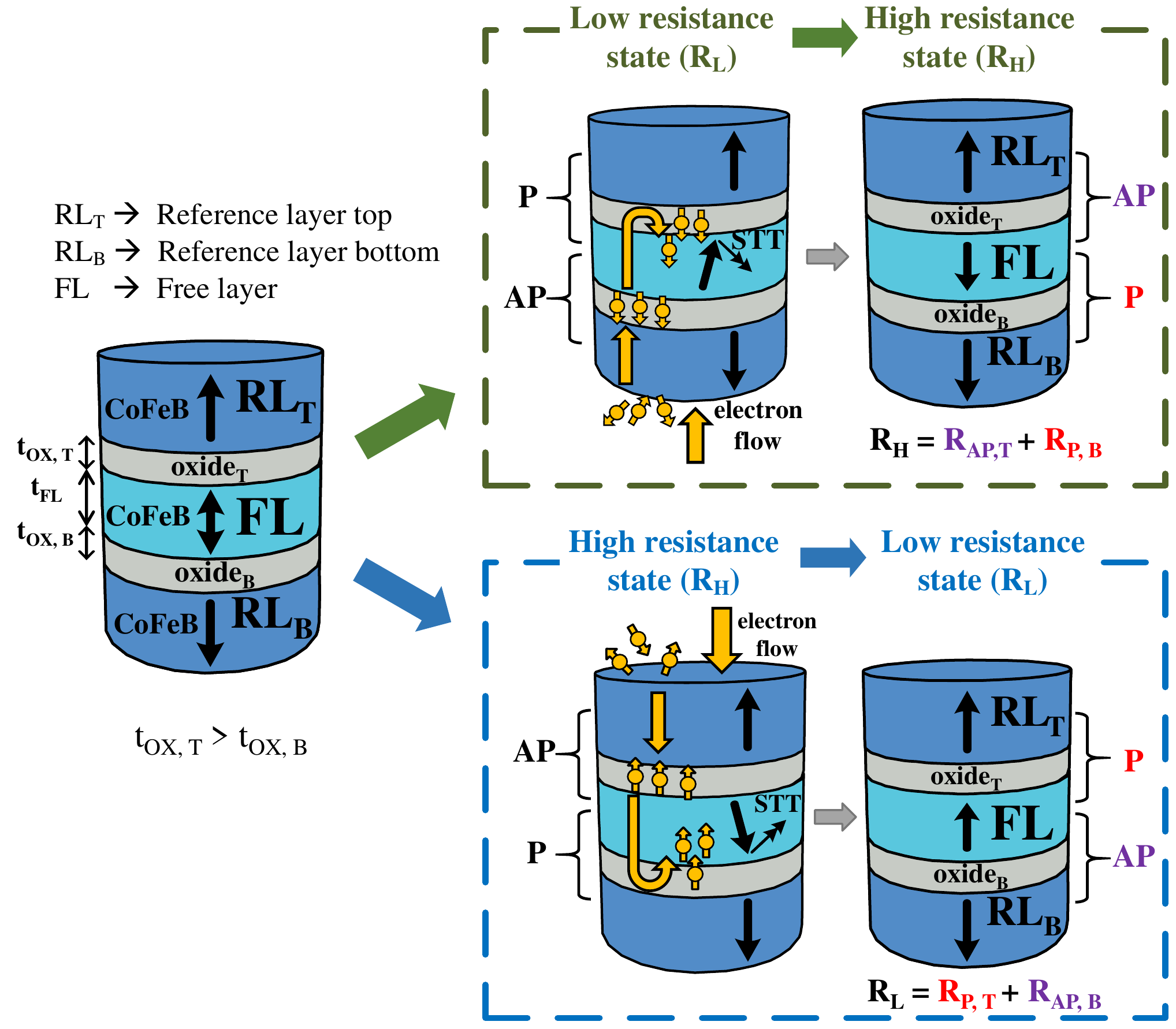}
    \vspace{-8mm}
    \caption{Sketch of the double-barrier magnetic tunnel junction (DMTJ) device along with STT switching description.}
    \label{fig:DMTJ}
\end{figure}

\begin{figure*}[!h] % [b]-> bottom, [t]->top, [H]->Here! ([h!] should do a better job), {figure*}->float
    \centering
    \includegraphics[width=1\columnwidth]{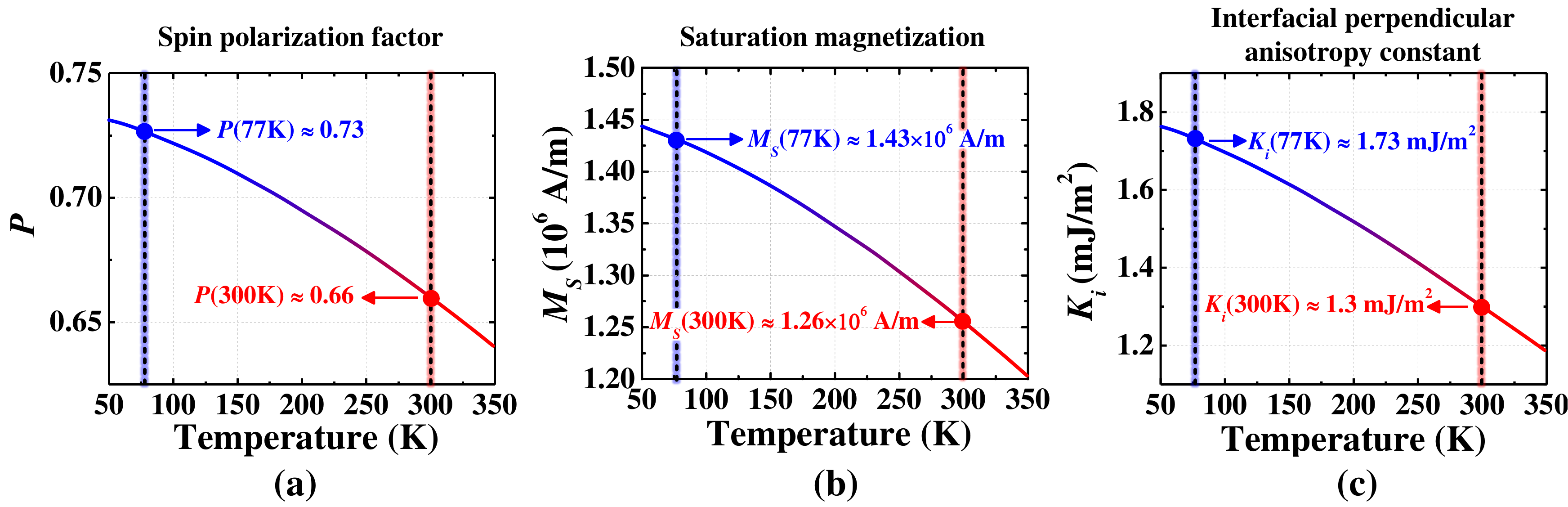}
    \vspace{-2mm}
    \caption{Temperature-dependent physical parameters of the DMTJ device: (a) spin polarization factor ($P$), (b) saturation magnetization (\Ms), and (c) interfacial perpendicular anisotropy constant (\Ki).}
    \label{fig:temp params}
\end{figure*}

\figref{fig:DMTJ} shows the structure of a perpendicular STT-\dmtj, which consists of a stack including three CoFeB \fm layers along with two thin MgO oxide barriers between the \fm layers~\cite{de2019compact}.
The outer \fm layers act as top and bottom polarizing reference layers (\RLtop and \RLbottom) with fixed magnetization orientation (antiparallel to each other). Conversely, the middle \fm layer, named \fl, has a variable magnetization orientation: parallel (P) or antiparallel (AP) with respect to that of the two RLs.
In addition, the two oxide layers feature different thicknesses, i.e., that of the top barrier (\toxt) is greater than the bottom barrier (\toxb). 
Overall, this implies two stable states associated with different resistances: low resistance (\RL) and high resistance (\RH) states (LRS and HRS).
More specifically, the LRS (HRS) corresponds to the \fl in P/AP (AP/P) configurations with respect to the \RLtop/\RLbottom.
Accordingly, \RL = \Rpt + \Rapb and \RH = \Rapt + \Rpb, where $R_{\text{P,T(B)}}$ and $R_{\text{AP,T(B)}}$ are the P and AP resistances associated with the top (bottom) barrier, respectively.
As shown in \figref{fig:DMTJ}, switching from one state to the opposite is carried out by the current-induced STT mechanism, i.e., by applying a current with appropriate direction through the device.
More precisely, the \RL$\rightarrow$\RH (\RH$\rightarrow$\RL) switching can be achieved by a current flowing from the \RLtop (\RLbottom) to the \RLbottom (\RLtop).
Thanks to the presence of the two RLs with opposite magnetization orientation, the DMTJ implies symmetric switching across the two transitions along with reduced switching currents as compared to the conventional \smtj~\cite{de2019compact,hu2015stt,hu2017low}.
This occurs at the cost of increased resistance and decreased tunneling magnetoresistance ($\mathit{TMR}$) ratio (as given by $\mathit{TMR}$ = (\RH - \RL) / \RL) owing to the extra oxide barrier~\cite{de2019compact,hu2015stt,hu2017low}.
However, the drawback of the reduced $\mathit{TMR}$ is alleviated when operating at cryogenic temperatures due to the intrinsic $\mathit{TMR}$ increase for lower temperatures, as discussed below. 

\begingroup
\setlength{\tabcolsep}{1.5pt}
\begin{table}[!b]
\centering
\begin{threeparttable}
\caption{DMTJ model parameters and characteristics.}
\footnotesize %\tiny, \scriptsize, \footnotesize, \small, \normalsize, \large, \Large, \LARGE, \huge, and \Huge
\begin{tabular}{c|c|cc}
\hline
\textbf{Parameter} & \textbf{Description}                                                                     & \multicolumn{1}{c|}{\textbf{\begin{tabular}[c]{@{}c@{}}DMTJ – 300\K\\    (40\nm, 13\nm)\end{tabular}}} & \textbf{\begin{tabular}[c]{@{}c@{}}DMTJ – 77\K\\    (40\nm, 13\nm)\end{tabular}} \\ \hline
\tfl               & FL thickness                                                                             & \multicolumn{2}{c}{1.2\nm}                                                                                                                                                              \\ \hline
\toxt              & Top barrier thickness                                                                    & \multicolumn{2}{c}{(1.0, 0.85)\nm}                                                                                                                                                      \\ \hline
\toxb               & Bottom barrier thickness                                                                 & \multicolumn{2}{c}{(0.5, 0.4)\nm}                                                                                                                                                       \\ \hline
$RA$                 & Resistance-area product                                                                  & \multicolumn{2}{c}{(7, 2) \ohms$\cdot$\umsquared}                                                                                                                                           \\ \hline
$P$                  & Spin polarization factor                                                                 & \multicolumn{1}{c|}{0.66}                                                                             & 0.73                                                                            \\ \hline
\Ms & Saturation magnetization & \multicolumn{1}{c|}{1.58\T}                                                                         & 1.80\T                                                                          \\ \cline{3-4} \hline
$\alpha$           & Gilbert damping factor                                                                   & \multicolumn{2}{c}{0.03~\cite{ikeda2010perpendicular,sampan2019temperature}}                                                                                                                                                                \\ \hline
\Ki                & \begin{tabular}[c]{@{}c@{}}Interfacial perpendicular \\ anisotropy constant\end{tabular} & \multicolumn{1}{c|}{1.3\X$10^{-3}$ \J/\msquared}                                                      & 1.73\X$10^{-3}$ \J/\msquared                                                    \\ \hline
$\mathit{TMR}(0)$             & \begin{tabular}[c]{@{}c@{}}Tunnel magnetoresistance\\ @0\V\end{tabular}                  & \multicolumn{1}{c|}{(141, 131)\%}                                                                     & (205, 198)\%                                                                    \\ \hline
\RL                & \begin{tabular}[c]{@{}c@{}}Low resistance\end{tabular}                 & \multicolumn{1}{c|}{(5.81, 17.8)\Kohms}                                                               & (5.99, 18.5)\Kohms                                                              \\ \hline
\RH                & \begin{tabular}[c]{@{}c@{}}High resistance\end{tabular}              & \multicolumn{1}{c|}{(14.0, 41.2)\Kohms}                                                               & (18.3, 55.2)\Kohms                                                              \\ \hline
\Ic               & Critical switching current                                                               & \multicolumn{1}{c|}{(13.6, --\tnote{*}~~)\uA}                                                                   & (16.6, 2.9)\uA                                                                  \\ \hline
$\Delta$           & Thermal stability factor                                                                 & \multicolumn{1}{c|}{(45, --\tnote{*}~~)}                                                                        & (175, 60)                                                                       \\ \hline
\end{tabular}
\label{tab:DMTJ parameters}
\begin{tablenotes}\footnotesize
\item[*]Note that \Ic and $\Delta$ is not reported here, since the 13\nm-DMTJ is not thermally stable at 300\K.
\end{tablenotes}
\end{threeparttable}
\end{table}
\endgroup

In this work, DMTJ operation is described by a macrospin-based Verilog-A compact model~\cite{de2019compact}, which includes temperature-dependence of physical parameters.
\tblref{tab:DMTJ parameters} reports the used parameters and resulting characteristics for two circular devices with different diameters (40\nm and 13\nm) at two operating temperatures (300\K and 77\K). 
300\K model parameters of \tblref{tab:DMTJ parameters}, such as resistance-area product ($RA$), spin polarization factor ($P$), saturation magnetization (\Ms), Gilbert damping factor ($\alpha$), and interfacial perpendicular anisotropy constant (\Ki), \hl{were calibrated according to experimental data at 300\,K reported in~\mbox{\cite{nowak2016dependence,sankey2008measurement,ikeda2010perpendicular}}.}
Note also that, in order to ensure full compatibility with the CMOS process, the resistance of the shrunk device is scaled down according to the trend reported in~\cite{apalkov2013sttmram}. 
\hl{Then, the temperature dependence of some parameters such as $P$, $M_S$ and $K_i$ was modeled by using semi-empirical laws widely used in literature~\mbox{\cite{kou2006temperature,zhao2015spintronics, zhang2018addressing,de2017compact}} as follows:} 
\begin{equation}
    \label{eqn:SpinPolarizationFactor}
     \SpinPolarizationFactor,  
\end{equation}
\begin{equation}
    \label{eqn:SaturationMagnetization}
     \SaturationMagnetization,   
\end{equation}
\begin{equation}
    \label{eqn:InterfacialPerpendicularAnisotropy}
     \InterfacialPerpendicularAnisotropy,   
\end{equation}
where the material-dependent fitting parameters $\beta$ and $T^{*}$ are respectively equal to $2\times10^{-5}\K^{-3/2}$~\cite{kou2006temperature} and $1120\K$~\cite{zhao2015spintronics}, whereas 0\K values (i.e., $P(0)$, $M_{S}(0)$, and $K_{i}(0)$) are set to meet 300\K values reported in \tblref{tab:DMTJ parameters}.
According to \eqref{eqn:SpinPolarizationFactor}-\eqref{eqn:InterfacialPerpendicularAnisotropy}, \figref{fig:temp params}(a)-(c) show the trend with the temperature of such parameters with emphasis on corresponding values at 300\K and 77\K.
As the temperature decreases, all three parameters increase, thus resulting in higher values at 77\K with respect to room temperature.
\hl{This is in line with experimental results reported in~\mbox{\cite{lang2020lowtemperature,rehm2019cryomtj}.}}

The physical parameters, mentioned above, affect  \dmtj characteristics such as resistance, $\mathit{TMR}$ ratio, critical switching current (\Ic), and thermal stability factor ($\Delta$), which in turn exhibit a temperature dependence as reported in \tblref{tab:DMTJ parameters}.
The temperature dependence of the resistance and $\mathit{TMR}$ ratio referred to each oxide barrier of the DMTJ is expressed by~\cite{kou2006temperature,zhang2018addressing}:
\begin{equation}
    \label{eqn:MTJconductance}
     \MTJconductance,  
\end{equation}
\begin{equation}
    \label{eqn:TMRzeroBias}
     \TMRzeroBias. 
\end{equation}
\begin{equation}
    \label{eqn:TMRvoltageDependent}
     \TMRvoltageDependent,   
\end{equation}
where $G(\theta)$ is the conductance in P ($\theta = 0$) or AP ($\theta = \pi$) state, $G_{T}$ and $G_{SI}$ are respectively the prefactor for direct elastic tunneling and the inelastic spin independent conductance term, $\mathit{TMR}(T,0)$ is the temperature-dependent $\mathit{TMR}$ at zero bias voltage, $\mathit{TMR}(T,V)$ is the $\mathit{TMR}$ including both temperature and voltage dependence, $V_\text{OX}$ is the voltage drop across the oxide barrier, and $V_{H}$ is the voltage drop for $\mathit{TMR}(T,V)$ = $\mathit{TMR}(T,0)$/2. 
From \eqref{eqn:MTJconductance}-\eqref{eqn:TMRzeroBias}, the $\mathit{TMR}$ of each barrier increases with decreasing temperature as given by the $P$ increase through \eqref{eqn:SpinPolarizationFactor}.
More specifically, the $\mathit{TMR}$ increase at lower temperatures is associated with an increase of the antiparallel resistance, whereas the parallel resistance is practically  temperature-independent\hl{, which is consistent with earlier experimental studies~\mbox{\cite{lang2020lowtemperature,cao2019lowtemp,rehm2019cryomtj}.}}
In the \dmtj of \figref{fig:DMTJ}, this implies increased $\mathit{TMR}$ and \RH at cryogenic temperatures, whereas \RL is roughly temperature-independent, as reported in \tblref{tab:DMTJ parameters}.

\hl{The $I_c$ of the DMTJ is modeled by~\mbox{\cite{de2019compact}}:}
\begin{equation}
    \label{eqn:MTJCriticalSwitchingCurrent}
     \MTJCriticalSwitchingCurrentVa,
\end{equation}
\hl{where $e$ is the electron charge, $\mu_{0}$ is the vacuum permeability, $\gamma_{0}$ is the absolute value of the gyromagnetic ratio, $V_\text{FL}$ is the volume of the FL, $\mu_{B}$ is the Bohr magneton, $k_{B}$ is the Boltzmann constant, and $g_\text{STT}$ refers to the the STT spin efficiency, which is expressed for the DMTJ as~\mbox{\cite{carpentieri2018micromagnetic,de2019compact}}:}
\begin{equation}
    \label{eqn:STTspinEfficiency}
     \gDMTJ.   
\end{equation}
\hl{The $H_\mathit{k,eff}$ is the effective anisotropy field as given by~\mbox{\cite{de2017compact}}:}
\begin{equation}
    \label{eqn:EffectiveAnisotropyField}
     \EffectiveAnisotropyField,   
\end{equation}
\hl{where $N_Z$ and $N_{X,Y}$ are respectively the effective demagnetizing factors in the perpendicular and in-plane directions, and $t_{FL}$ is the FL thickness.\\
The $\Delta$ is modeled by~\mbox{\cite{sato2014properties,wang2019compact}}:}
\begin{equation}
    \label{eqn:deltaDW}
     \deltaDW,   
\end{equation}
\hl{where, $D$ is the device diameter, $D_W = \pi\sqrt{A_{ex}/K_{EFF}}$ is the domain wall width, $A_{ex}$ = 20\,pJ/m is the exchange stiffness constant, and $K_{EFF}$ is the effective PMA energy density as given by}
\begin{equation}
    \label{eqn:KEFF}
     \KEFFa. 
\end{equation}

\hl{In line with the above modeling, for a given temperature, $\Delta$ is proportional to the anisotropy and volume of the FL when $D$ is smaller than $D_W$ (i.e., when single-domain magnetization reversal takes place), whereas $\Delta$ does not scale with the device area when $D > D_W$.
Conversely, the critical switching current $I_c$ exhibits monotonic decrease with the decrease of $D$, as reported in~\mbox{\cite{sato2014properties,wang2019compact}}. 
As a consequence, the $\Delta/I_c$ ratio monotonically increases with decreasing $D$ until $D > D_W$, whereas it becomes constant for $D \leq D_W$.}
According to \eqref{eqn:MTJCriticalSwitchingCurrent}-\eqref{eqn:KEFF}, under cryogenic operation, the \dmtj exhibits higher $\Delta$ (i.e., better data retention capability) at the cost of higher switching currents (see \tblref{tab:DMTJ parameters}), as determined by the increase in \Ms and \Ki with decreasing temperature through \eqref{eqn:SaturationMagnetization}-\eqref{eqn:InterfacialPerpendicularAnisotropy}.

\begin{figure}[!t] % [b]-> bottom, [t]->top, [H]->Here! ([h!] should do a better job), {figure*}->float
    \centering
    \includegraphics[width=0.6\columnwidth]{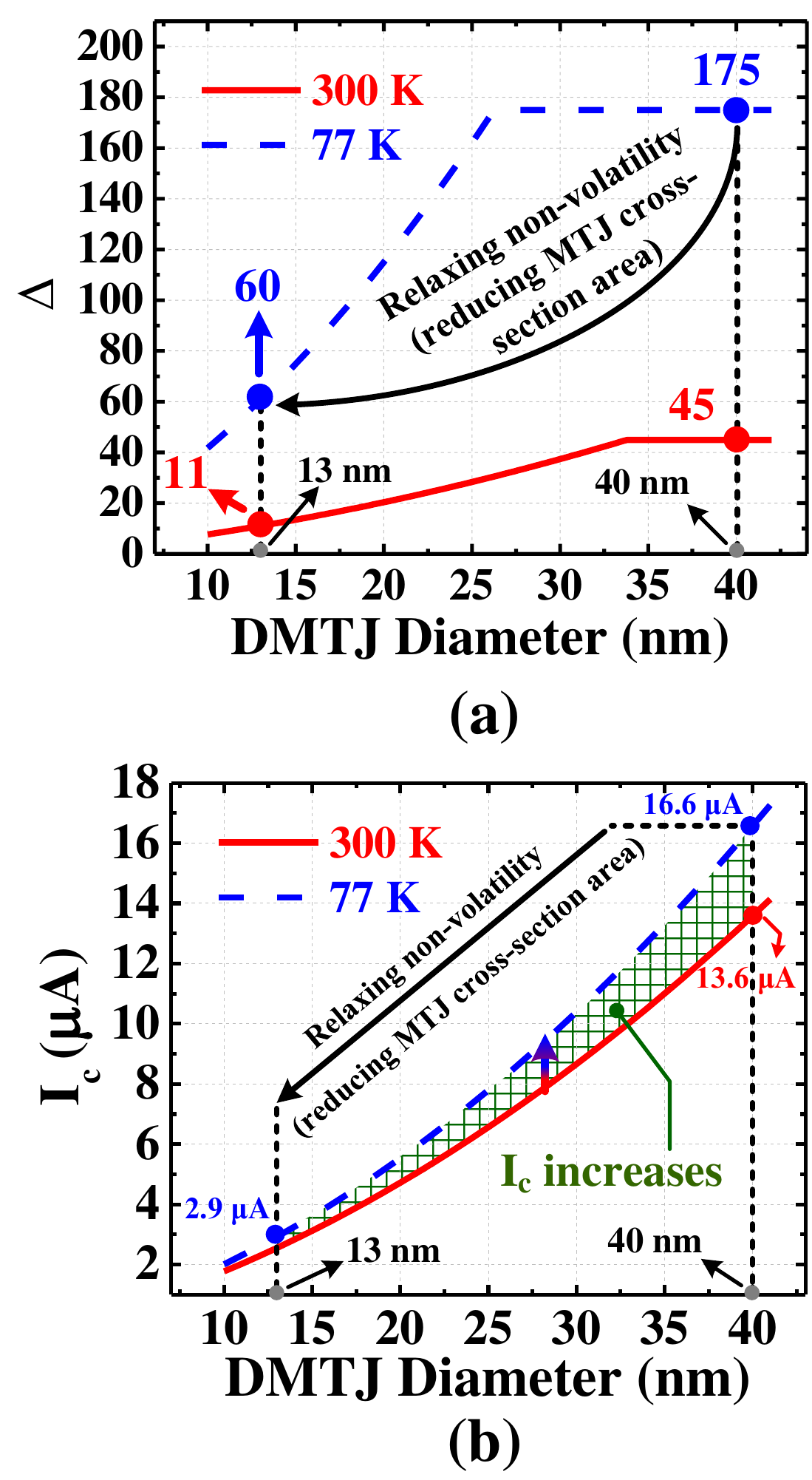}
    \vspace{-2mm}
    \caption{Temperature-dependent characteristics of the DMTJ device as function of the cell diameter: (a) thermal stability factor ($\Delta$) and (b) critical switching current (\Ic).}
    \label{fig:Ic0_delta_DMTJ}
    \vspace{-3mm}
\end{figure}

\begin{figure}[t!] % [b]-> bottom, [t]->top, [H]->Here! ([h!] should do a better job), {figure*}->float
    \centering
    \includegraphics[width=0.6\columnwidth]{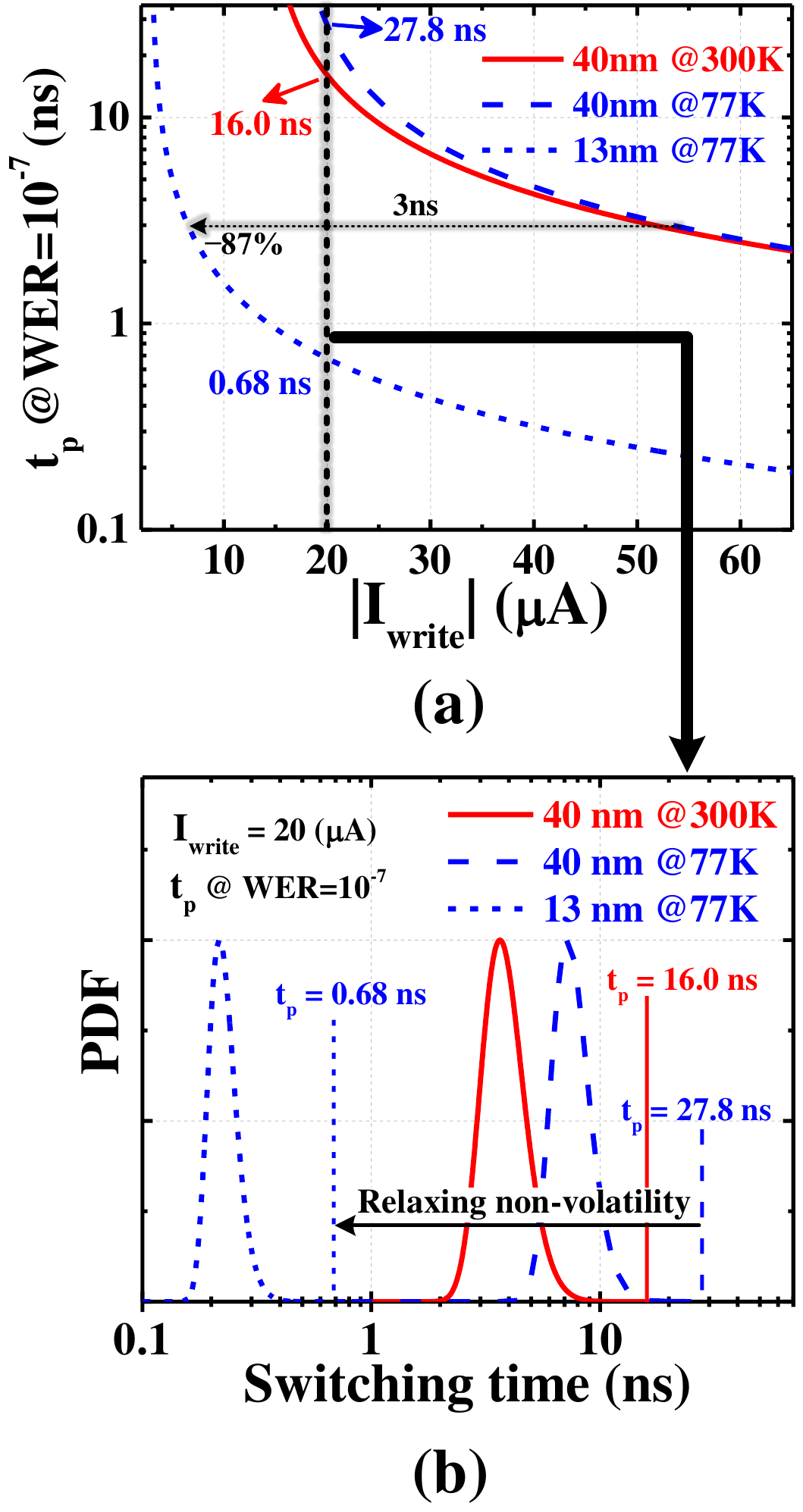}% 1 5
    \vspace{-3mm}
    \caption{Switching behavior of the DMTJ devices with different diameters (40\nm and 13\nm) at 300\K and 77\K: (a) write pulse width (\tp) to ensure a write error rate (WER) of $10^{-7}$ as a function of write current (\Iwrite), and (b) probability distribution function (PDF) of the switching time for \Iwrite = 20\uA.
    Data for the 13\nm-DMTJ at 300\K is not reported since it is thermally unstable at room temperature according to $\Delta$ values of \figref{fig:Ic0_delta_DMTJ}(a).}
    \label{fig:switching_DMTJ}\vspace{-4mm}
\end{figure}

The above discussed tradeoff between data retention capability and switching current can be effectively faced by properly tuning the thermal stability of \dmtj devices operating at cryogenic conditions~\cite{chiang2021cold,garzon2021relaxing}.
In other words, the idea is to significantly relax the non-volatility requirement at room temperature by reducing the cross-section area, while meeting the typical 10-years retention time at cryogenic temperatures.
In this regard, \figref{fig:Ic0_delta_DMTJ}(a) and (b) show the temperature-dependent $\Delta$ and \Ic, respectively, as function of the \dmtj diameter.
From \figref{fig:Ic0_delta_DMTJ}(a), the \dmtj with 40\nm diameter exhibits $\Delta\approx45$ at 300\K.
According to  \eqref{eqn:MTJCriticalSwitchingCurrent}-\eqref{eqn:deltaDW}, the same device shows $\Delta\approx175$ at 77\K, which corresponds to a significant increase in the \Ic by about 20\% (see \figref{fig:Ic0_delta_DMTJ}(b)) with a detrimental effect on energy consumption under write access. 
As shown in \figref{fig:Ic0_delta_DMTJ}(a)-(b), decreasing the cell diameter down to 13\nm allows maintaining 10-years data retention (i.e., $\Delta\approx60$) at 77\K, while enabling significantly lower \Ic.
The latter thus results in considerably lower energy consumption during the write operation, which is particularly sought after for embedded memories operating at cryogenic temperatures~\cite{garzon2022embedded}.
This can be appreciated in \figref{fig:switching_DMTJ}(a)-(b), which report the switching characteristics of the DMTJ with different diameters (i.e., 40\nm and 13\nm) at 300\K and 77\K.
From \figref{fig:switching_DMTJ}(a), the 13\nm DMTJ allows reducing the write current (\Iwrite) to achieve a target write error rate (WER) of $10^{-7}$~\cite{garzon2019exploiting} at 77\K for a given write pulse width (\tp) of 3\ns by more than 85\% as compared to the 40\nm device.
At the same time, for a given \Iwrite = 20\uA, the \tp required for WER = $10^{-7}$ at 77\K is reduced from 27.8\ns down to 0.68\ns when decreasing the cell diameter from 40\nm down to 13\nm, as highlighted in \figref{fig:switching_DMTJ}(b). 

\begin{figure}[b!] % [b]-> bottom, [t]->top, [H]->Here! ([h!] should do a better job), {figure*}->float
    \centering \vspace{-5mm}
    \includegraphics[width=1\columnwidth]{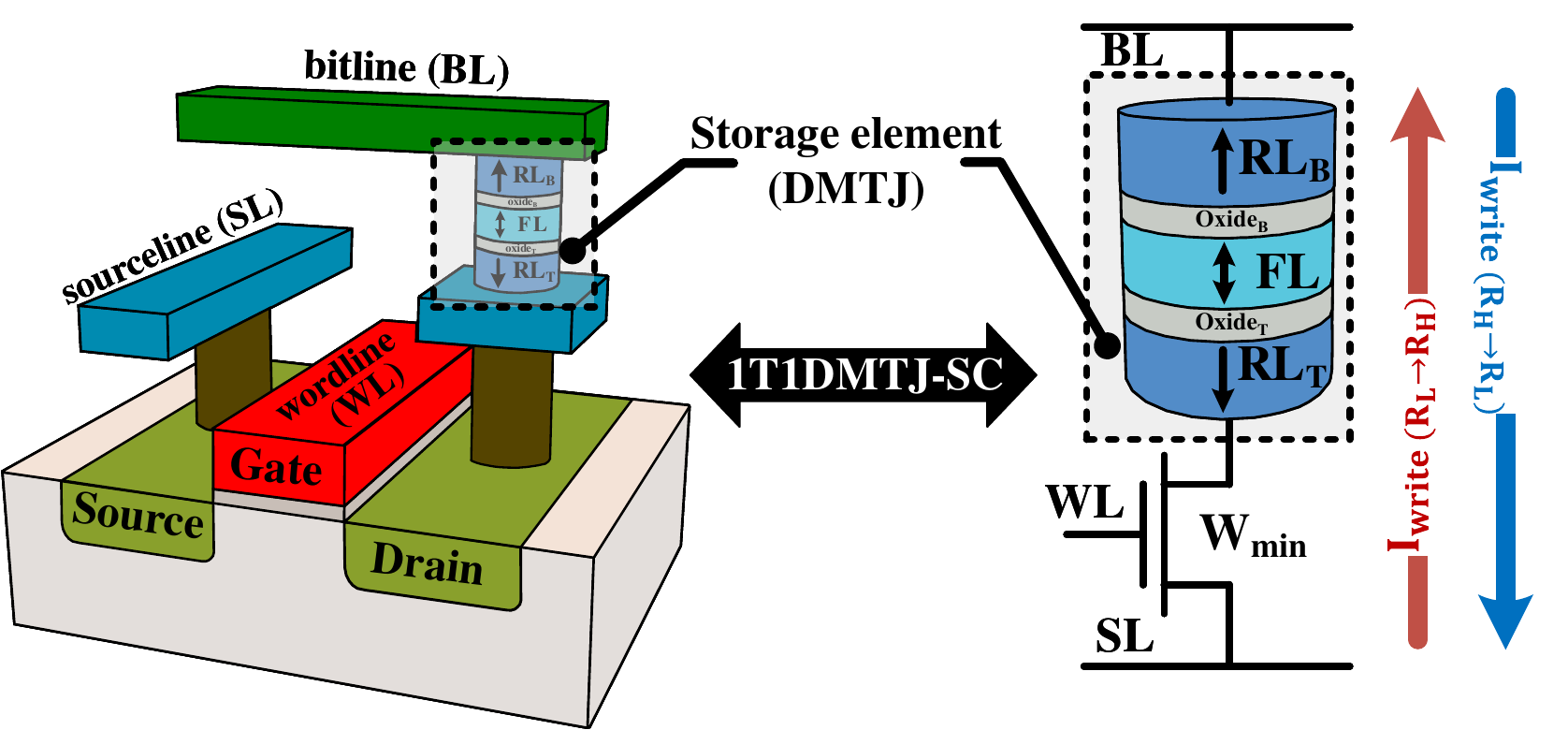}
    \vspace{-4mm}
    \caption{Schematic of the DMTJ-based STT-MRAM bitcell including the access transistor and the storage element: one transistor-one DMTJ in standard connection (1T1DMTJ-SC).}
    \label{fig:sttmram_cell}
    %\vspace{-2mm}
\end{figure}

\section{Analysis of DMTJ-based STT-MRAM at 77\K}
\label{sec:Analysis77K}
\subsection{Bitcell-level simulation results}

\figref{fig:sttmram_cell} shows the schematic of the considered \sttmram bitcell, which includes the access transistor and the storage element, typically placed between the second and fourth metal layers (M2 and M4)~\cite{zhao2013synchronous}.
Among the different DMTJ-based bitcell topologies, here we refer to the one transistor-one \dmtj in standard connection (i.e., 1T1DMTJ-SC where the minimum-sized access transistor is connected to the \RLtop), which allows the best trade-off between area and energy~\cite{garzon2019exploiting,garzon2021ultralow}.

Bitcell-level analysis was performed within the Cadence Virtuoso environment using the Spectre simulator, while exploiting transistor models of the commercial 65\nm CMOS technology and the Verilog-A based \dmtj compact model described in the previous section.
Extensive \mc simulations were carried out at the operating temperature of 77\K, while considering the effect of both \dmtj and CMOS process variability.
For the DMTJ device, Gaussian-distributed variations were considered with a variability (\var) of 5\% for the cross-section area and 1\% for \toxt, \toxb, and \tfl, in line with state-of-the-art literature~\cite{ho2013physics,ohashi2017variability,  sakhare2020j, de2019compact, de2016variability,Trinh2016voltage}.

\begin{figure}[!t] % [b]-> bottom, [t]->top, [H]->Here! ([h!] should do a better job), {figure*}->float
    \centering
    \includegraphics[width=1\columnwidth]{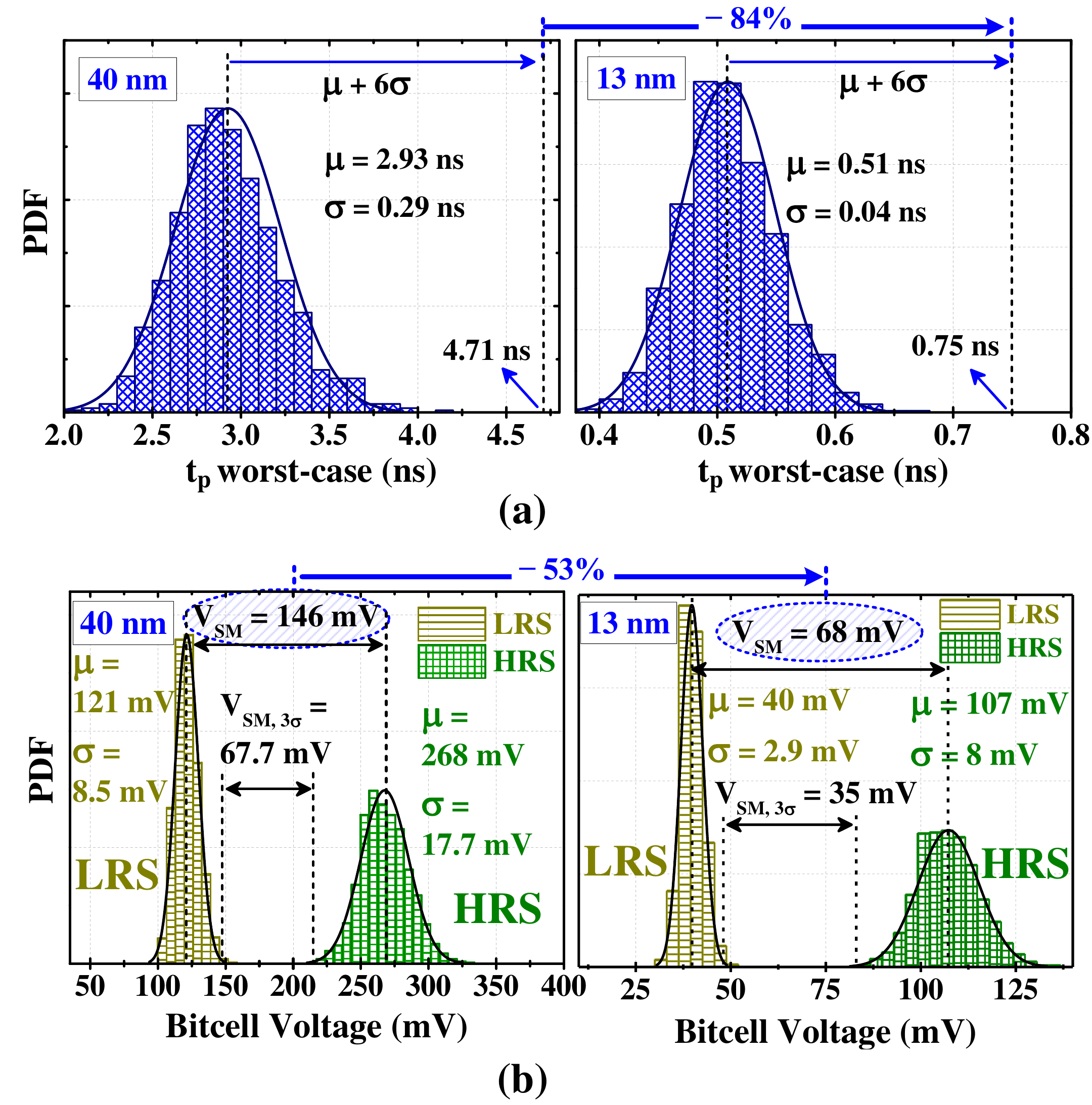}
    \vspace{-7mm}
    \caption{STT-MRAM bitcell-level results at 77\K for the DMTJ devices with different diameters (40\nm and 13\nm). (a) Write operation: statistical distribution of the worst-case \tp for WER=$10^{-7}$. (b) Read operation: statistical distribution of the bitcell voltage for a read current (\Iread) to ensure a RDR=$10^{-9}$.}
    \label{fig:sttmram_circuit_level}
    \vspace{-5mm}
\end{figure}

\figref{fig:sttmram_circuit_level} shows the bitcell-level \mc results
under both (a) write and (b) read operations when considering the \dmtjs with different diameters (40\nm and 13\nm) at 77\K. 
More precisely, \figref{fig:sttmram_circuit_level}(a) reports the statistical distribution of the \tp that ensures the target WER of $10^{-7}$ in reference to the worst-case switching transition.
From this figure, thanks to the reduced \Ic, the 13\nm \dmtj-based bitcell allows reducing the \tp evaluated at the 6$\sigma$ corner by $\sim$84\% as compared to its 40\nm \dmtj-based counterpart.
\figref{fig:sttmram_circuit_level}(b) shows comparative results in terms of the statistical distribution of the bitcell voltage referred to the read operation. 
The latter is carried out using a conventional voltage sensing scheme~\cite{garzon2019exploiting}, i.e., by applying a fixed read current (\Iread) through the bitcell and comparing the bitcell voltage with a reference voltage by a sense amplifier.   
The \Iread has to be set sufficiently lower than the \Ic to ensure a reasonably low \rdr, i.e., the probability of unintentionally switching the stored data during the read operation.
Here, we considered a target \rdr of $10^{-9}$ along with a read pulse width (\tread) of 1\ns~\cite{quang2016boosted}. This results in \Iread equal to 15.6\uA and 1.91\uA for the 40\nm and 13\nm \dmtj-based bitcells, respectively.
\figref{fig:sttmram_circuit_level}(b) also highlights the voltage sensing margin (\Vsm), i.e., the difference between bitcell voltages corresponding to the DMTJ in HRS and LRS, estimated at both nominal and $3\sigma$ corner.
From this figure, the 13\nm \dmtj-based bitcell shows smaller nominal ($3\sigma$) \Vsm by $\sim$53\% ($\sim$48\%) when compared to its 40\nm \dmtj-based counterpart.
This is ascribed to the reduced \Ic and hence \Iread, despite the increased resistance of the 13\nm device (see \tblref{tab:DMTJ parameters}).

\begin{table}[!b] %b
\centering
\vspace{-5mm}
\caption{STT-MRAM bitcell-level results at 77\K for write and read operations.}
\small %\tiny, \scriptsize, \footnotesize, \small, \normalsize, \large, \Large, \LARGE, \huge, and \Huge
\begin{tabular}{c|cc}
\hline
\multirow{2}{*}{\textbf{Write operation (@WER  = $10^{-7}$)}} & \multicolumn{2}{c}{\textbf{DMTJ-based bitcell}}      \\ \cline{2-3} 
                                                              & \multicolumn{1}{c|}{\textbf{40\nm}} & \textbf{13\nm} \\ \hline
\RL$\rightarrow$\RH \Iwrite (\uA)                               & \multicolumn{1}{c|}{54.9}           & 26.0           \\
\RH$\rightarrow$\RL \Iwrite (\uA)                               & \multicolumn{1}{c|}{92.6}           & 44.8           \\
Worst-case 6$\sigma$ \tp (\ns)                                  & \multicolumn{1}{c|}{4.71}           & 0.75           \\
Average write energy (\fJ)                                    & \multicolumn{1}{c|}{393}            & 33.0           \\ \hline
\multirow{2}{*}{\textbf{Read operation (@RDR = $10^{-9}$)}}   & \multicolumn{2}{c}{\textbf{DMTJ-based bitcell}}      \\ \cline{2-3} 
                                                              & \multicolumn{1}{c|}{\textbf{40\nm}} & \textbf{13\nm} \\ \hline
\Iread (\uA)                                            & \multicolumn{1}{c|}{14.6}           & 1.91           \\
Read power (\pW)                                             & \multicolumn{1}{c|}{17.5}           & 2.3           \\
LRS resistance (\Kohms)                                       & \multicolumn{1}{c|}{8.3}           & 20.8           \\
HRS resistance (\Kohms)                                       & \multicolumn{1}{c|}{18.3}           & 56.1           \\
Nominal \Vsm (\mV)                                  & \multicolumn{1}{c|}{146.3}            & 68.0             \\ 
3$\sigma$ \Vsm (\mV)                                  & \multicolumn{1}{c|}{67.7}            & 35.0             \\ 
\hline
\end{tabular}
\label{tab:summary_cells}
\end{table}

Finally, \tblref{tab:summary_cells} summarizes bitcell-level simulation results for write and read operations at 77\K. 
Reported data shows that the 13\nm \dmtj-based bitcell is more energy-efficient under both write/read access (--92\%/--87\%), while also exhibiting better write performance as compared to its 40\nm \dmtj-based counterpart.
These advantages are obtained at the cost of reduced (about halved) sensing margin during the read operation.
However, this drawback can be properly addressed as proposed in~\cite{Trinh2018dynamic}.

\subsection{Architecture-level simulation results}
\begin{figure}[t!] % [b]-> bottom, [t]->top, [H]->Here! ([h!] should do a better job), {figure*}->float
    \centering
    \includegraphics[width=0.6\columnwidth]{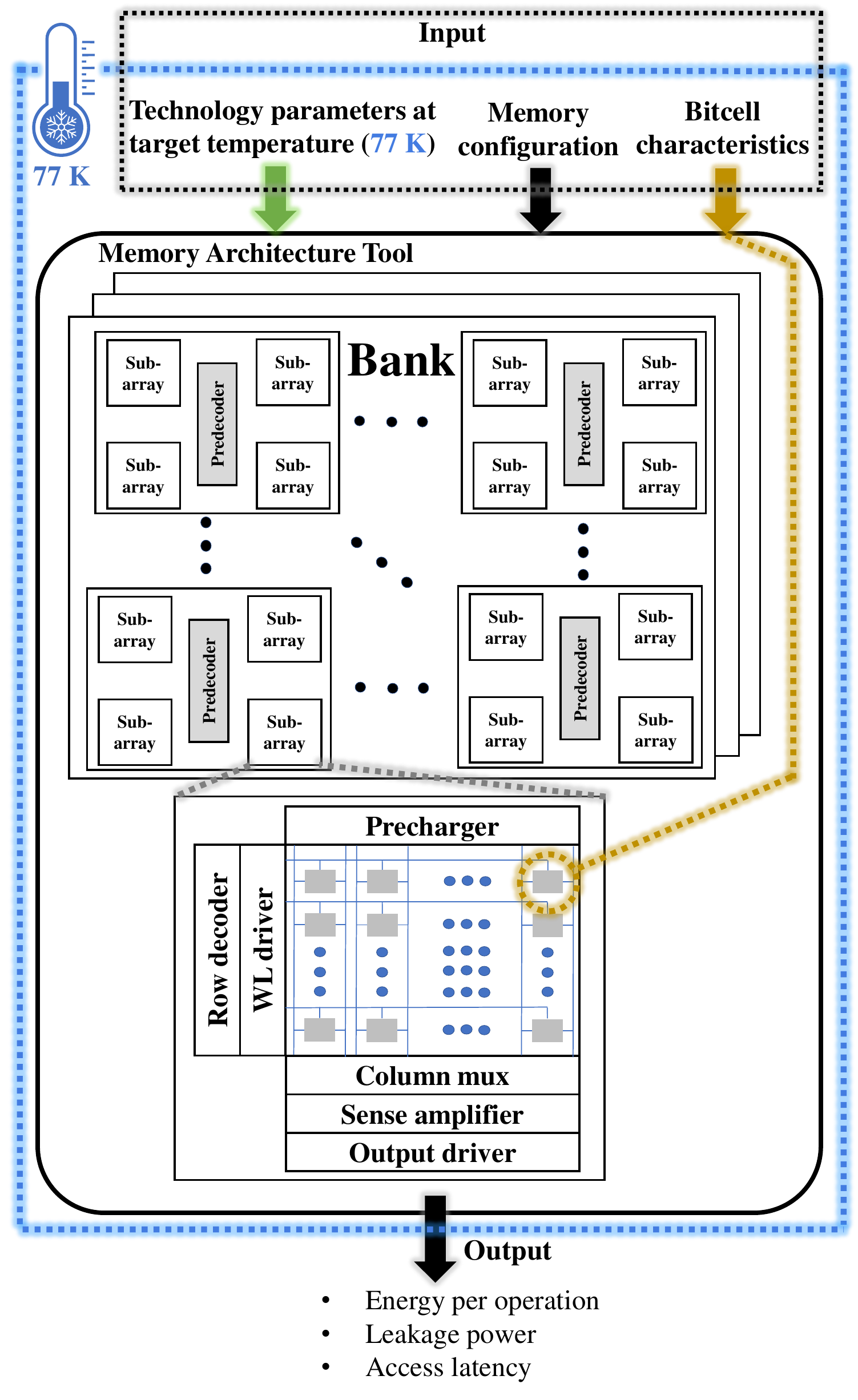}
    %\vspace{-8mm}
    \caption{Modeling approach at the memory architecture-level within the DESTINY estimation tool~\cite{poremba2015destiny}.}
    \label{fig:DEST_tool_framework}
    %\vspace{-4mm}
\end{figure}

The memory architecture-level analysis was carried out using the DESTINY estimation tool~\cite{poremba2015destiny} to evaluate the main \sttmram figures of merit, such as access latency, energy per operation and leakage power.
\figref{fig:DEST_tool_framework} shows the overview of the modeling approach used in the DESTINY tool, where the memory bank architecture is organized as an array of sub-blocks, called Mats.
Each Mat consists of multiple sub-arrays along with a predecoder.
The sub-array represents the core memory block, which includes the bitcell array and some peripheral circuits needed to manage read/write operations.

For the sake of accuracy and to keep consistency with the above bitcell-level analysis, the DESTINY tool was calibrated for an operating temperature of 77\K, as shown in \figref{fig:DEST_tool_framework}.
The read/write electrical characteristics reported in \tblref{tab:summary_cells} were used to describe the bitcell within the memory sub-array, while also specifying information about its area and aspect ratio along with the width of the access transistor.
In addition, temperature-dependent technology parameters of the adopted 65\nm CMOS process including both transistor (e.g., threshold voltage, ON/OFF current, mobility, gate capacitance, etc.) and interconnection (e.g., metal resistivity) characteristics were embedded into the source code of DESTINY for an accurate modeling of peripheral memory circuits.

\begin{figure}[t!] % [b]-> bottom, [t]->top, [H]->Here! ([h!] should do a better job), {figure*}->float
    \centering
    \includegraphics[width=0.6\columnwidth]{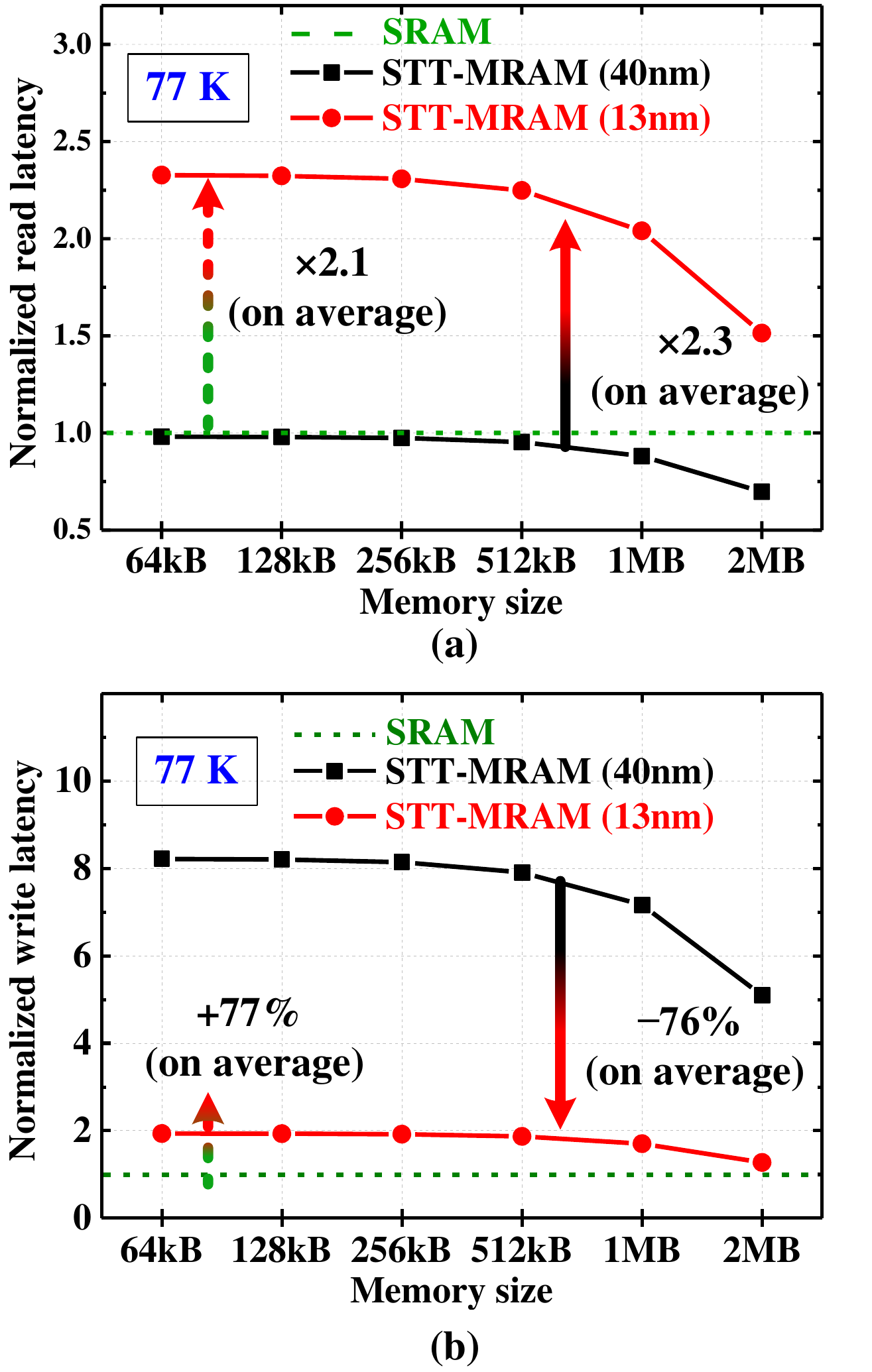}
    %\vspace{-6mm}
    \caption{Architecture-level results at 77\K of 40\nm and 13\nm DMTJ-based STT-MRAMs versus 6T-SRAM in terms of (a) read and (b) write latency for memory sizes ranging from 64\KB up to 2\MB . Data is normalized to conventional 6T-SRAM.}
    \label{fig:latency_results}
    \vspace{-4mm}
\end{figure}

\begin{figure}[t!] % [b]-> bottom, [t]->top, [H]->Here! ([h!] should do a better job), {figure*}->float
    \centering
    \includegraphics[width=0.6\columnwidth]{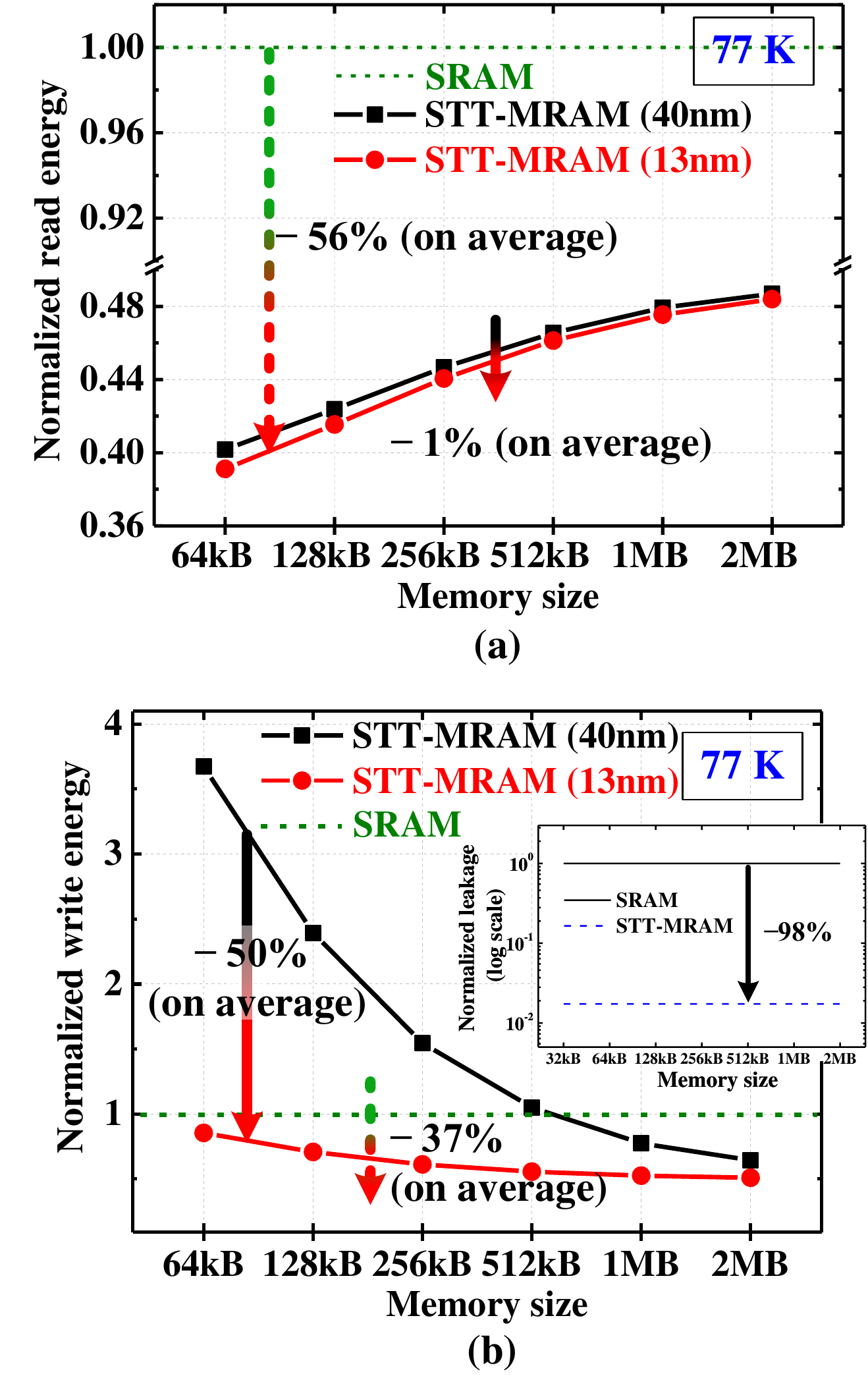}
    %\vspace{-6mm}
    \caption{Architecture-level results at 77\K of 40\nm and 13\nm DMTJ-based STT-MRAMs versus 6T-SRAM in terms of (a) read and (b) write dynamic energy consumption for memory sizes ranging from 64\KB up to 2\MB . Leakage power in the inset. Data is normalized to conventional 6T-SRAM.}
    \label{fig:energy_results}
    \vspace{-4mm}
\end{figure}

At the memory architecture-level, \sttmrams based on \dmtjs with 40\nm and 13\nm diameter were benchmarked against conventional \sixtsram for different cache sizes, ranging from 64\KB up to 2\MB.
\figref{fig:latency_results}(a)-(b) and \figref{fig:energy_results}(a)-(b) show obtained DMTJ-based STT-MRAM results under read/write access in terms of operation latency and dynamic energy consumption, where data is normalized to \sixtsram. 
Leakage power is also reported in the inset of \figref{fig:energy_results}(b).
From \figref{fig:latency_results}(a), the STT-MRAM based on 13\nm \dmtj devices exhibits increased read latency by 2.3\X and 2.1\X (averaged over the considered memory capacity range) as compared to the 40\nm DMTJ-based STT-MRAM and the \sixtsram, respectively.
\hl{This is mainly due to the lower \mbox{\Iread} applied during read operation (refer to Table 2), which arises from the reduced $I_c$ of the 13\,nm device (see Fig. 4(b)).}
From \figref{fig:latency_results}(b), both DMTJ-based STT-MRAMs exhibit a performance penalty under write access with respect to the \sixtsram.
However, the use of 13\nm DMTJ devices with relaxed retention time allows significantly reducing this penalty.
Indeed, the 13\nm DMTJ-based STT-MRAM decreases write latency by more than 75\%, on average, as compared to its 40\nm DMTJ-based counterpart.
Note that, as compared to \sixtsram, the \sttmram exhibits a read/write access time penalty, although it is reduced as cache memory capacity increases~\cite{garzon2019exploiting}. 
As shown in \figref{fig:energy_results}(a)-(b), using 13\nm \dmtj devices also allows the STT-MRAM to outperform the \sixtsram in terms of dynamic energy consumption under both read/write accesses.
From \figref{fig:energy_results}(a), the two STT-MRAM implementations show comparable read energy with a reduction of more than 50\% (56\% for the 13\nm DMTJ-based STT-MRAM) on average than the \sixtsram.
In addition, from \figref{fig:energy_results}(b), the 13\nm DMTJ-based STT-MRAM halves write energy as compared to the 40\nm DMTJ-based memory, thus ensuring energy savings of --37\% on average with respect to the \sixtsram.
Looking at the leakage power shown in the inset of \figref{fig:energy_results}(b), the two STT-MRAM implementations exhibit similar static power consumption, which is 98\% lower than the \sixtsram. 
Indeed, although the leakage is considerably reduced by orders of magnitude at cryogenic temperatures~\cite{garzon2021gain}, \sixtsram implementations still suffers from significant static power, mainly consumed by the bitcell array~\cite{garzon2021cryoSB,garzon2021cryoDB}.

\section{Conclusions}
\label{sec:Conclusions}
In this work, a solution to build reliable, energy-efficient, and high-density \sttmrams operating at cryogenic conditions (77\K) has been proven.
It consists of using \dmtj devices to exploit their reduced switching currents, while relaxing their non-volatility requirement at room temperature (i.e., by reducing the cross-section area) and maintaining the typical 10-years retention time at 77\K.
Our simulation study has been performed at bitcell-level and memory architecture-level by using a commercial 65\nm CMOS technology calibrated down to 77\K under silicon measurements and a macrospin-based \dmtj Verilog-A compact model. 
As the main contribution of our work, obtained results have shown that shrinking the DMTJ cell size (i.e., relaxing its retention time at room temperature) allows improved performance and energy consumption under write access at 77\K. 
This makes \dmtj-based \sttmram operating at 77\K more energy-efficient for both read/write operations than conventional \sixtsram, at the only cost of worsened read access time.

\section*{Acknowledgment} 
This work was partially supported by the Israel Science Foundation under Grant 996/18, by the Smart Imaging Consortium under the MAGNET program of the Israel Innovation Authority, and by the project PRIN 2020LWPKH7 funded by the Italian Ministry of University and Research.   

\bibliography{abbreviations_long,bibfile}
\biboptions{sort&compress} % to compress when multiple refs are used [1,2,3,4] -> [1-4]

\end{document}